\documentclass[11pt]{article}
\usepackage{graphicx}
\usepackage[utf8]{inputenc}
\usepackage{lscape}
\usepackage{fullpage}
\usepackage{setspace}
\usepackage{amssymb}
\usepackage{url}
\usepackage{xcolor}
\usepackage{parskip}
\usepackage{float}
\usepackage[section]{placeins}
\usepackage[square,numbers]{natbib}
\begin{document}
\doublespacing
\title{visTree: Visualization of Subgroups for a Decision Tree}
\author{A. Venkatasubramaniam$^\mathsection$ and J. Wolfson$^\dagger$\\
\textit{$^\mathsection$The Alan Turing Institute, British Library, London, UK}\\
\textit{$^\dagger$Division of Biostatistics, School of Public Health, University of Minnesota, Minneapolis, USA}}
\maketitle

\abstract{
Decision trees are flexible prediction models which are constructed to quantify outcome-covariate relationships and characterize relevant population subgroups. However, the standard graphical representation of fitted decision trees highlights individual split points, and hence is suboptimal for visualizing defined subgroups. In this paper, we present a novel visual representation of decision trees which shifts the primary focus to characterizing subgroups, both in terms of their defining covariates and their outcome distribution. We implement our method in the \texttt{visTree} package, which builds on the toolkit and infrastructure provided by the \texttt{partykit} package and enables the visualization to be applied to varied decision trees. Individual functions are demonstrated using data from the Box Lunch study [French et al., 2014], a randomized trial to evaluate the effect of exposure to different lunch sizes on energy intake and body weight among working adults.}

\section{Introduction}

\begin{figure}[h!]
\centering
\includegraphics[scale = 0.32]{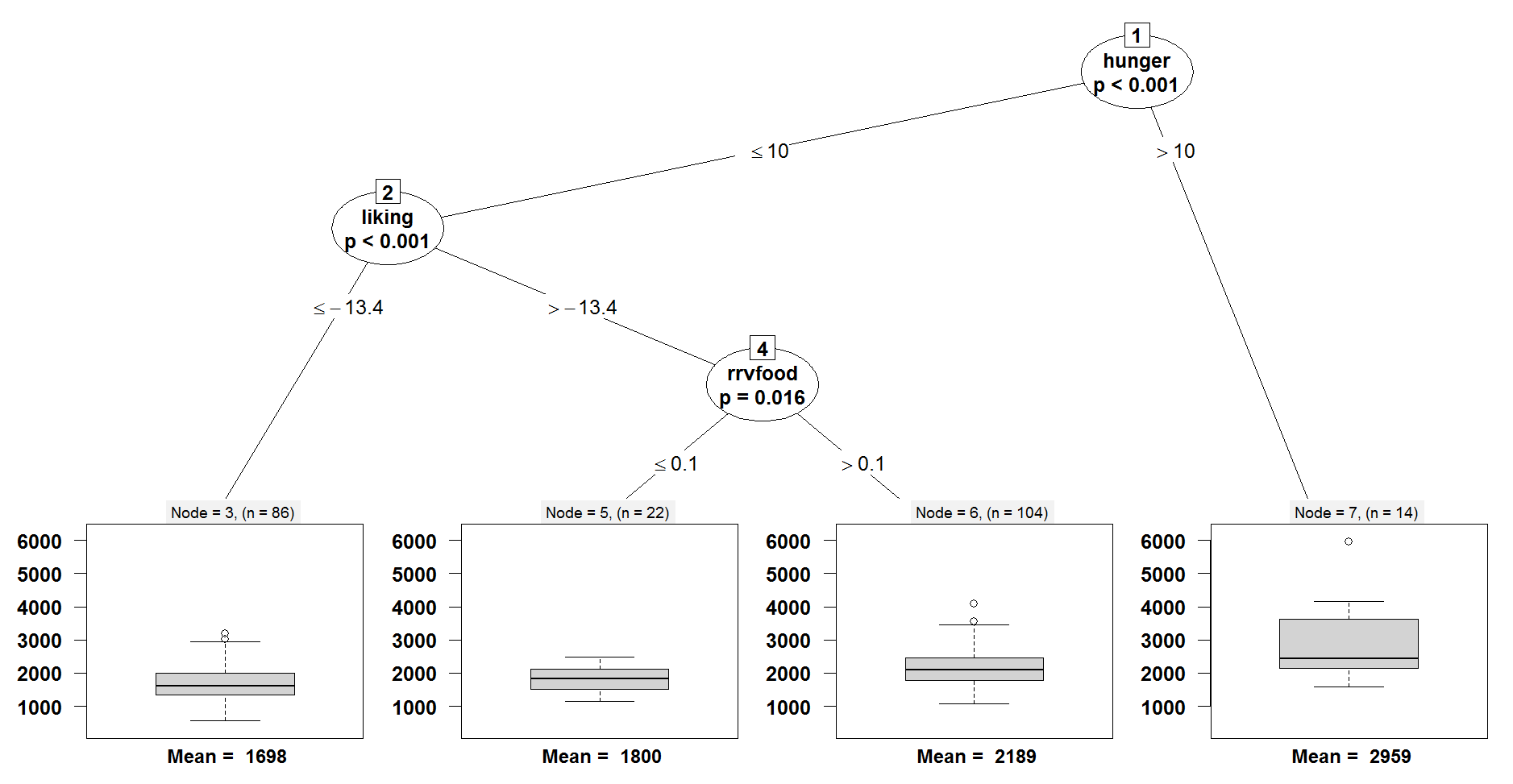}
\caption{Conditional inference tree that represents the relationship between six eating behavior covariates (\textbf{hunger}, \textbf{wanting}, \textbf{liking}, relative reinforcement of food (\textbf{rrvfood}), \textbf{disinhibition}, and restrained eating (\textbf{rest\_eating})) and daily caloric intake in kilo-calories (\textbf{kcal24h0}).}\label{fig:tree}
\end{figure}

Machine learning techniques \citep{Cortes1995,Ripley2007,Friedman1996} provide flexible modeling of covariate-outcome relationships, but they are “black box” techniques which often present limited insight about the nature of these relationships. Recursive partitioning provides a simple and systematic approach to building flexible prediction models which can be interpreted in a straightforward manner. Beyond the ability to determine predictions, recursive partitioning also creates a graph-based structure called decision tree which depicts a hierarchical partitioning of the data. Decision trees identify distinct sub-populations by selecting covariates associated with an outcome of interest and are relatively straightforward to construct. Subgroups identified by decision trees encourage decision making that target distinct sub-populations, where each subgroup is composed of a set of outcome values that correspond to defined characteristics. For example, we utilize data from the Box lunch study (BLS), which is a randomized controlled trial designed to evaluate effects of portion sizes on calorie intake and weight gain \citep{Simone2014, French20142}. The data is composed of 226 observations recorded for 25 relevant covariates and an outcome variable and we seek to identify subgroups of individuals that correspond to varying daily caloric intake levels. In Figure \ref{fig:tree}, a decision tree represents a relationship between Energy in kilo-calories consumed through the day and six eating behavior covariates (\textbf{hunger}, \textbf{liking}, \textbf{wanting}, relative reinforcement of food (\textbf{rrvfood}), \textbf{disinhibition}, and restrained eating (\textbf{rest\_eating})). Each terminal node corresponds to a distinct sub-population of calorie intake levels defined by individual characteristics.

\noindent One popular method of constructing trees is the so-called Classification and Regression Tree (CART) algorithm \citep{Breiman1984}. CART and related techniques (e.g., C4.5 \citep{Quinlan2014}) have seen wide application, but they also have some drawbacks. For example, the exhaustive search approach used by the CART algorithm has a tendency to split the sample on variables with multiple possible splits. If one binary and one continuous-valued variable each have a similar association with the outcome, CART is more likely to split the sample on the continuous-valued variable. This is commonly referred to as the biased variable selection problem \citep{Loh2002}. Further, CART is extremely greedy in searching for optimal splits, which makes it prone to over-fitting so that careful pruning is required to obtain a tree which predicts well in independent data \citep{Esposito1997}. An alternative to CART is the Conditional Inference Tree (CTree) technique proposed by \cite{Hothorn2006}. Based on theory developed by \cite{Strasser1999}, CTree performs recursive partitioning using a sequence of statistical hypothesis tests. By constructing trees in this way, CTree avoids the problem of biased variable selection and is easily generalized to a variety of covariate and outcome types including categorical, continuous, ordinal, and time-to-event. Further, CTrees provide an intuitive way of controlling model complexity and over-fitting by setting a p-value threshold. Existing packages within R implement decision tree methods such as the \texttt{rpart} \citep{Therneau1997} package for CART, \texttt{rpart.plot} \citep{Milborrow2011} for plotting \texttt{rpart} trees, the \texttt{partykit} \citep{Hothorn2015} package  for CTree, the \texttt{RWeka} \citep{hornik2009} package for pruned or unpruned \texttt{C4.5}, and the \texttt{C50} \citep{Kuhn2014} package for an extension of the \texttt{C4.5} classification algorithms. 
The \texttt{partykit} package \citep{Hothorn2015} provides a common infrastructure for recursive partitioning; independent of the decision tree method and its underlying software implementation. Ensemble models over decision trees (such as random forests) typically generate results with a higher predictive accuracy and a reduced tendency to overfit training data but limit the user's ability to discern a single set of decision rules. \\

\noindent In general, the ability to derive meaningful conclusions from decision trees is dependent on an understanding of the response variable and their relationship with associated covariates identified by splits at each node of the tree. However, the standard tree view makes it challenging to characterize these subgroups. This is for two reasons: 1) the covariate values at splitting points do not necessarily indicate whether the groups are balanced or imbalanced, and 2) When a particular covariate is split on multiple times, it can be hard to keep track of which range of covariate values defines the resulting subgroup. For example, a researcher unfamiliar with the range of values for \textbf{hunger}, \textbf{liking}, and \textbf{rrvf} may struggle to understand and interpret the subgroups defined by the CTree in Figure \ref{fig:tree}: is the group defined by \textbf{hunger} $> 10$ a relatively extreme group consisting of those with very high \textbf{hunger}, or does it also include individuals with moderate \textbf{hunger} levels? \\

\noindent In this paper, we introduce a package \texttt{visTree} that provides a graphical visualization for characterizing individual subgroups identified by a decision tree. The package \texttt{visTree} utilizes tools provided within the framework of the \texttt{partykit} package and we explore its implementation in R using relevant examples. We introduce individual components of the \texttt{visTree} package, provide an interpretation of the visualization and illustrate applications to multiple scenarios. 
In addition, this visualization can also be applied to other decision trees such as CART by utilizing appropriate coercion functions from the \texttt{partykit} package. \texttt{visTree} package is available as a stable version on CRAN and for development purposes at \texttt{https://github.com/AshwiniKV/visTree}.


\section{Conditional Inference Trees}
\label{sect:ctree}

Conditional inference trees introduced by \cite{Hothorn2006} recursively partition the sample data into mutually exclusive subgroups that are maximally distinct with respect to a defined parameter (e.g., the mean). The primary idea of the conditional inference tree is that determining the variable to split on and the optimal split on the selected variable are performed as sequential steps rather than simultaneously. 
The first step determines whether to split the tree, and if so, which variable to split on. Let $\mathbf{y} = (y_1, \dots, y_n)$ be a response vector with accompanying $m$-dimensional predictor vector $\mathbf{X} = \{X_1 \dots X_m\}$. The decision of whether or not to split is based on an evaluation of the global null hypothesis, $H_0: P(\mathbf{y}|\mathbf{X}) = P(\mathbf{y})$. Since $H_0 = \displaystyle{\cap_{j=1}^m} H_0^j$, the global null hypothesis is tested by considering the set of partial null hypotheses $\{H_0^j, j = 1, \dots, m\}$. A p-value is computed for each $H_0^j$ and when the minimum of the computed p-values is lower than $\alpha$, the global null hypothesis is rejected and the tree should be split. Otherwise, if $H_0$ is not rejected, no more splits are made and the node is declared a terminal node. The threshold for the minimum p-value is determined by applying a multiple comparison adjustment (e.g., Bonferroni) to a fixed level of significance, $\alpha$. If splitting is to be done, the variable to split on is chosen by considering the t-statistic corresponding to the null hypothesis $H_0^j: \beta_j = 0$ in the univariate linear regression $E(\mathbf{y} |X_j) =  \beta_0+ \beta_j X_j$ (this same t-statistic is also used to compute the needed p-value for $H_0^j$). 


Once a covariate (say $X_l$) has been selected for splitting, the actual split is chosen such that the discrepancy between the groups is maximized. 
In Figure \ref{fig:tree}, the tree selects the covariate \textbf{hunger} and partitions the population into two groups: 1) subjects whose \textbf{hunger} measurement is less than or equal to 10, and 2) subjects whose \textbf{hunger} is greater than 10. Each terminal node displays a box plot that represents the distribution of 24-hour energy intake for the subjects contained in the node. The prediction for an individual observation from the training set in the node is typically obtained as the node-specific sample mean. For example, the terminal node ID = 3 in Figure \ref{fig:tree} contains subjects that have a 24 hour mean energy intake of 1698, where this intake value is displayed below the terminal node and these subjects have characteristics represented by a tree pathway defined as \textbf{hunger} $\leq$ 10 and \textbf{liking} $\leq $ -13.4. 
The ability to assign a parameter value that determines the size of the CTree provides an alternative to pruning and overcomes the biased variable selection inherent in CART. 

\subsection{partykit package}

The \texttt{partykit} \citep{Hothorn2015} package provides a formal and `unified' infrastructure for recursive partitioning in R and additional tools for the purpose of summarizing, representing, and generating predictions from decision trees. The \texttt{ctree()} function in \texttt{partykit} implements the conditional inference tree with output generated as a `party' object. For example, a conditional inference tree can be constructed to investigate the relationship between daily energy intake (\textbf{kcal24h0}) and a number of covariates in the BLS data using:

\begin{verbatim}
library(partykit)
ptree<-ctree(kcal24h0 ~ hunger + wanting + liking + disinhibition +
			    resteating + rrvfood, data = blsdata)
\end{verbatim}

\noindent A \texttt{`party'} object generated by \texttt{ctree()} is a recursive partition structure and built using splits of class \texttt{`partysplit'} and nodes of class \texttt{`partynode'}. \texttt{`partynode'} is a class that typically represents inner nodes and terminal nodes and \texttt{`partysplit'} is a class that represents splits at inner nodes. 
An option \texttt{ctree\_control} within \texttt{ctree()} allows the user to make adjustments in the tree building process and includes arguments to modify aspects such as the minimum value of the defined test statistic that must be exceeded in order to allow a covariate to split (\texttt{mincriterion}), the minimum sum of weights in a terminal node (\texttt{minbucket}), the minimum sum of weights within a node for the node to be considered for splitting (\texttt{minsplit}), the maximum depth of the tree (\texttt{maxdepth}), etc.  
In general, the toolkit provided by \texttt{partykit} allows for individual functions to be easily modified and extended and we use this infrastructure to build the \texttt{visTree} package. In the next section, we introduce the \texttt{visTree} package and demonstrate its ability to implement a visualization for characterizing subgroups identified by a decision tree. 

\section{visTree package}
\label{sect:visTree}

\begin{figure}[ht]
\centering
\includegraphics[scale = 0.4]{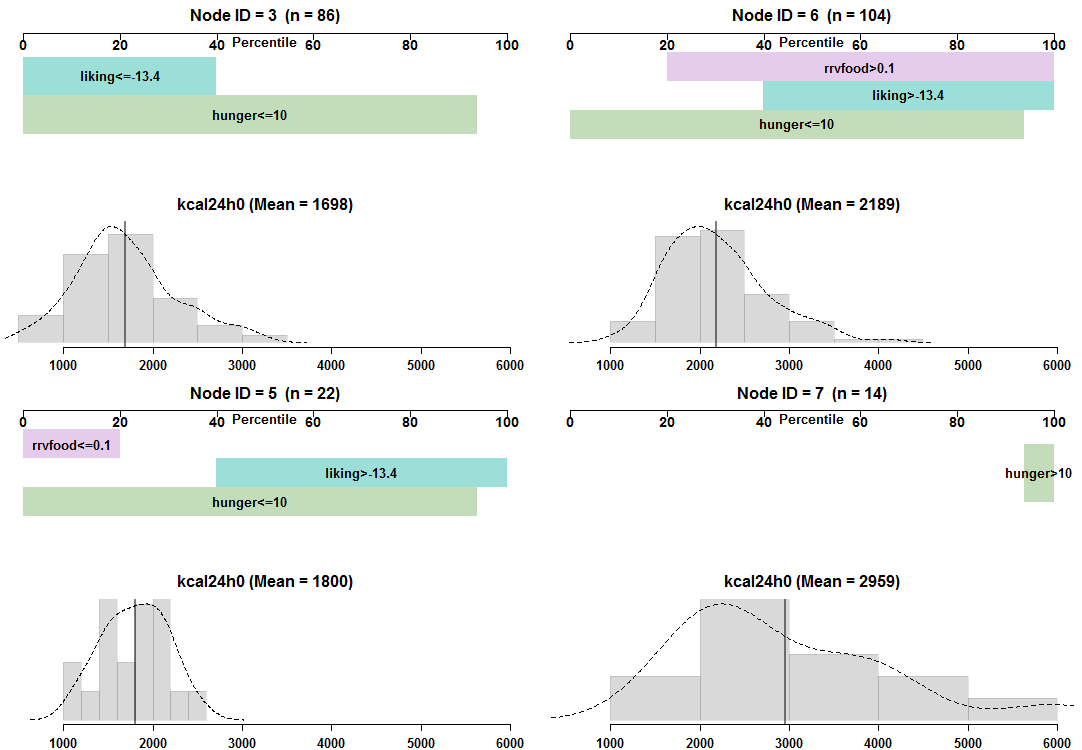}
\caption{The \texttt{visTree()} function is applied to the conditional inference tree in Figure \ref{fig:tree} to visualize and enable subgroup characterization.}\label{fig:vistree}
\end{figure}

The \texttt{visTree} package provides a new way of visualizing decision trees that makes subgroup characterization and interpretation more transparent. Visualizations such as tree maps and sectioned scatter plots have been previously developed \citep{Urbanek2002,Urbanek2008} using the classical view of a  tree-based model. For example, \cite{Urbanek2008} provides representations that describe proportions using colors, such that they are placed within the inner and terminal nodes of the tree. Figure \ref{fig:vistree} shows the result of applying the \texttt{visTree()} function to the conditional inference tree displayed in Figure \ref{fig:tree}, where Figure \ref{fig:tree} predicts daily caloric intake as a function of individual personality characteristics. 

\begin{verbatim}
visTree(ptree, text.axis = 1.3, text.label = 1.2, text.bar = 1.2, 
        alpha = 0.5, text.percentile = 1.2)
\end{verbatim}

\noindent Each subplot in Figure \ref{fig:vistree} is a self-contained summary of one of the subgroups defined by the terminal nodes in Figure \ref{fig:tree} (Node ID = 3, 5, 6, and 7). The subplot title gives the mean daily caloric intake and number of individuals within the subgroup. The histogram in the background shows the distribution of the outcome for individuals in this group; the mean of observations within each bin defining the histogram is displayed at the bottom of each bar, and the overall mean is indicated by a vertical line. The key information in each subplot, however, is contained in the colored horizontal bars which summarize the set of covariate splits used to define the subgroup. For example, the top left sub-plot of Figure \ref{fig:vistree} corresponds to the terminal node with node ID = 3 ($n = $ 86) in Figure \ref{fig:tree}.  This particular subgroup is composed of individuals with low to moderate values of \textbf{liking} (i.e., $\leq$ -13.4, approximately the 40th population percentile) and all but the highest \textbf{hunger} values (i.e., $\leq$ 10, approximately the 92nd population percentile). The overall mean daily caloric intake within this group is 1698 kcal. Subplots for terminal nodes ID = 5 and ID = 6 represent subgroups defined by the same splits for \textbf{liking} and \textbf{hunger}, but with different values for \textbf{rrvfood}; node ID = 5 corresponds to a group with low \textbf{rrvfood} ($\leq$ 0.11, the 20th population percentile), while node ID = 6 is the complementary group with \textbf{rrvfood} above the 20th population percentile). The subplot for terminal node ID = 7 represents a subgroup defined by \textbf{hunger} $>$ 10 (i.e., very high \textbf{hunger}), with this group having a very high daily energy intake in comparison to the other defined subgroups. 

\subsection{Multiple splits}

\begin{figure}[ht]
\centering
\includegraphics[scale = 0.25]{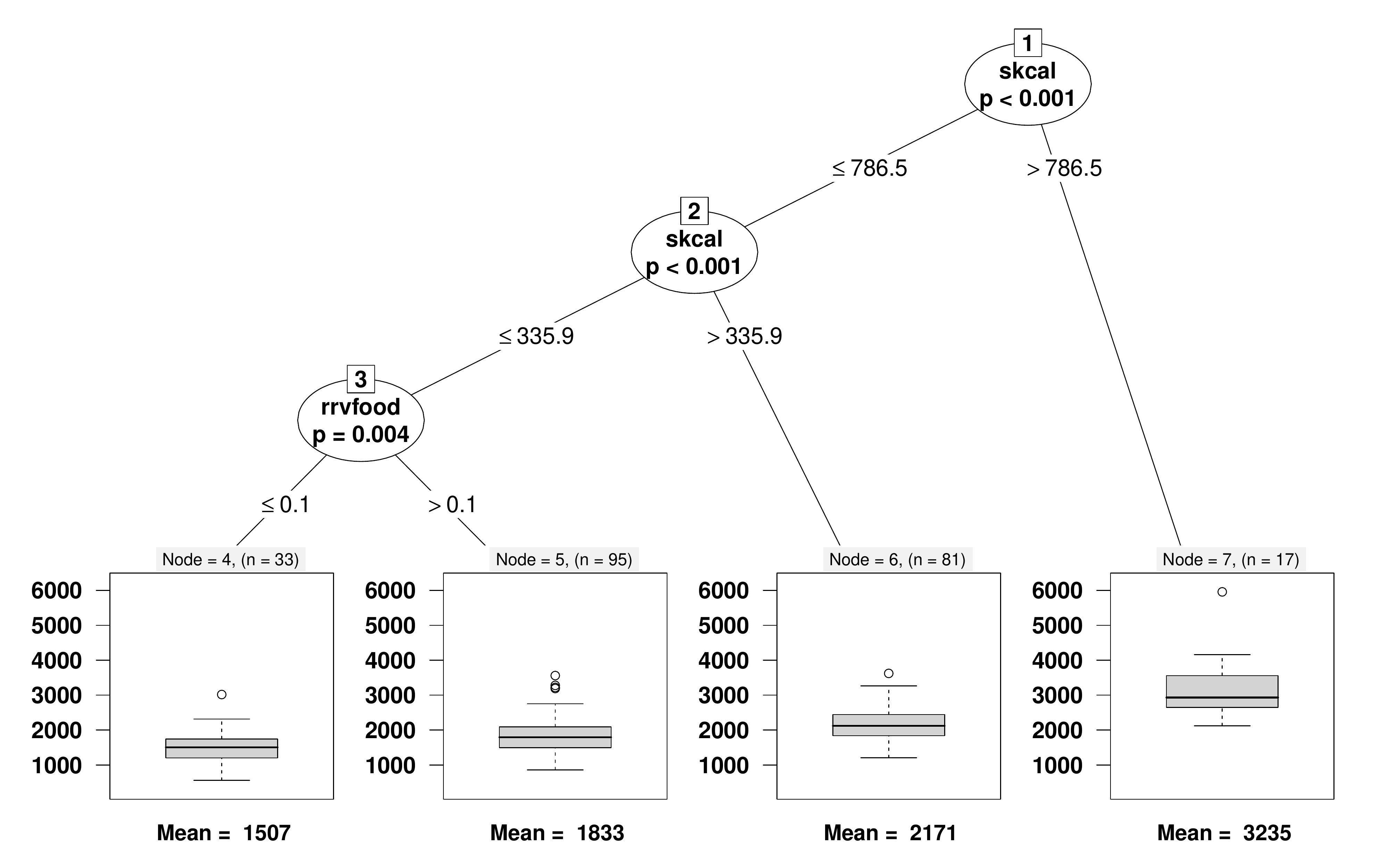}
\caption{Decision tree with multiple splits on the same covariate.}\label{fig:reptree}
\end{figure}

\begin{figure}[ht]
\centering
\includegraphics[scale = 0.4]{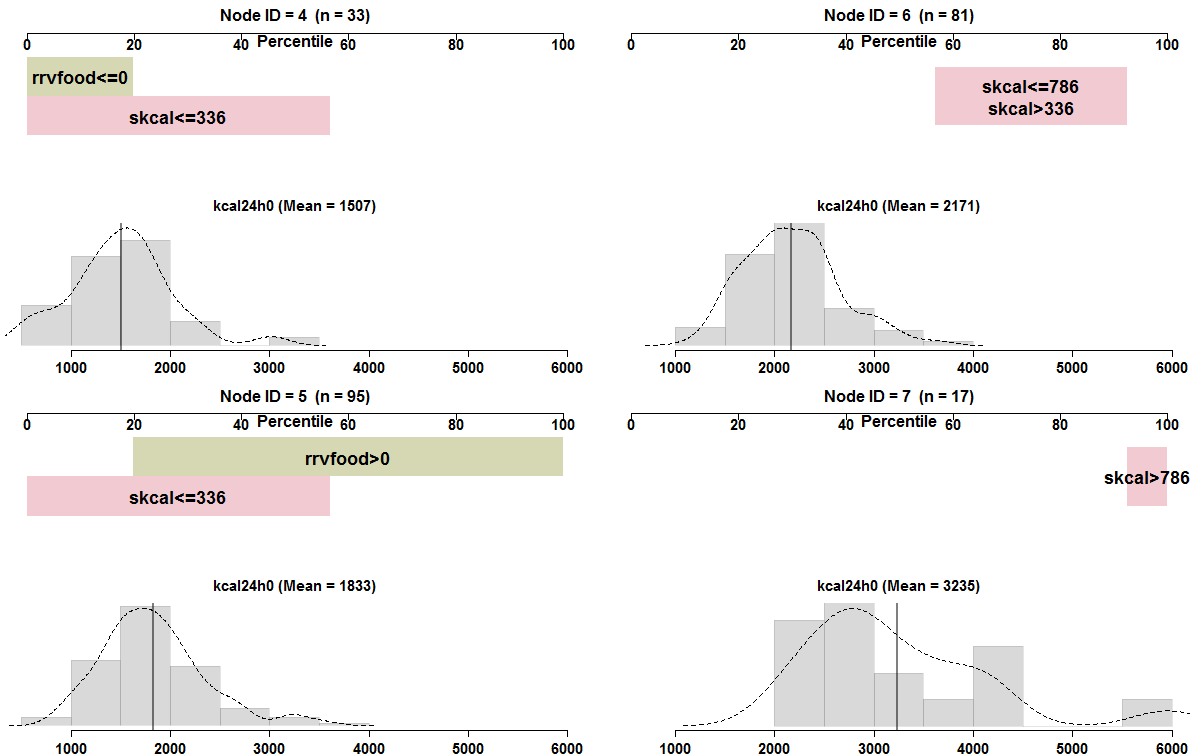}
\caption{The \texttt{visTree} function consolidates repeated splits on \textbf{skcal} in Figure \ref{fig:reptree} to display intervals which define subgroups.}\label{fig:visrep}
\end{figure}

Decision trees may split on the same covariate multiple times, in which case it can be particularly challenging to summarize and interpret subgroups using the standard branch-and-node tree visualization. Consider, for example, Figure \ref{fig:reptree}, where snack kilo-calories (\textbf{skcal}) is split on at the root node and at node ID = 2. The corresponding \texttt{visTree} visualization is shown in Figure \ref{fig:visrep}; splits on \textbf{skcal} are consolidated to display the intervals of \textbf{skcal} defining the splits (see node ID = 6). 

\begin{verbatim}
ptree2 <- ctree(kcal24h0 ~ skcal + rrvfood + resteating + age, data = blsdata)
plot(ptree2)            
visTree(ptree2, alpha = 0.5, text.round = 0, text.main = 1.4, text.axis = 1.4, 
        text.bar = 1.75, text.label =1.4, text.percentile = 1.5)
\end{verbatim}

\subsection{Categorical outcomes}

\begin{figure}[ht]
\centering
\includegraphics[scale = 0.225]{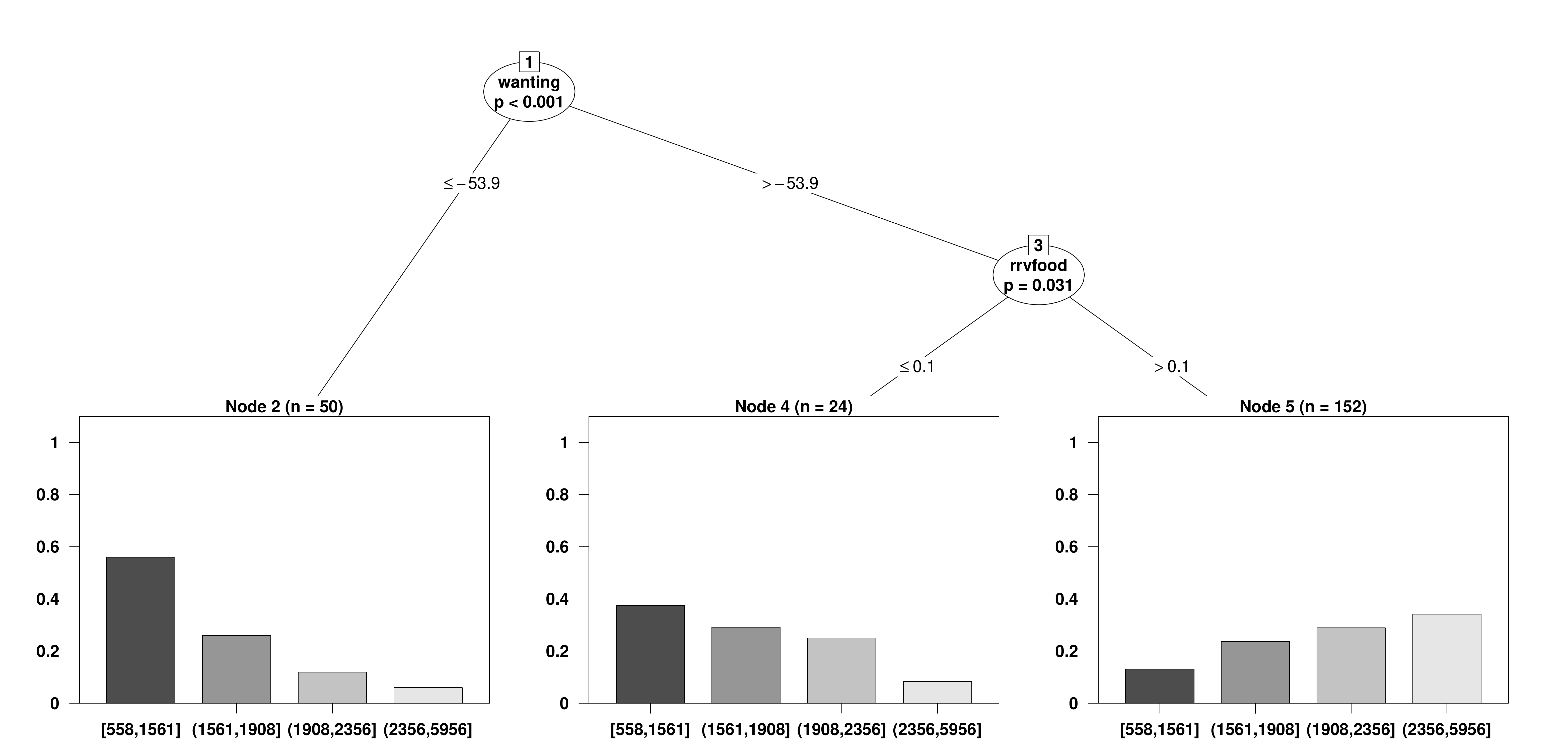}
\caption{Conditional inference tree that represents a  relationship between Energy in kilo-calories treated as a categorical response variable and six defined continuous covariates.}\label{fig:catctree}
\end{figure}

Decision trees can also be used to represent the relationships between covariates and a categorical outcome. In Figure \ref{fig:catctree}, a CTree is used to display the covariates and a categorical outcome defined by partitioning daily caloric intake into four categories: $\left\{[558, 1561], (1561, 1908], (1908, 2356], (2356, 5956]\right\}$. The terminal node displays a bar chart to describe each category of the response variable and the corresponding conditional class probabilities in the identified subgroup. 

\begin{verbatim}
blsdataedit<-blsdata[,-7]
blsdataedit$bin<-0
blsdataedit$bin<-cut(blsdata$kcal24h0, unique(quantile(blsdata$kcal24h0)),
					  include.lowest = TRUE, dig.lab = 4)
names(blsdataedit)[26]<-"kcal24h0"
ptree3<-ctree(kcal24h0 ~ hunger + rrvfood + resteating + liking + 
		             wanting + disinhibition, data = blsdataedit)
plot(ptree3)
\end{verbatim}

\begin{figure}[ht]
\centering
\includegraphics[scale = 0.4]{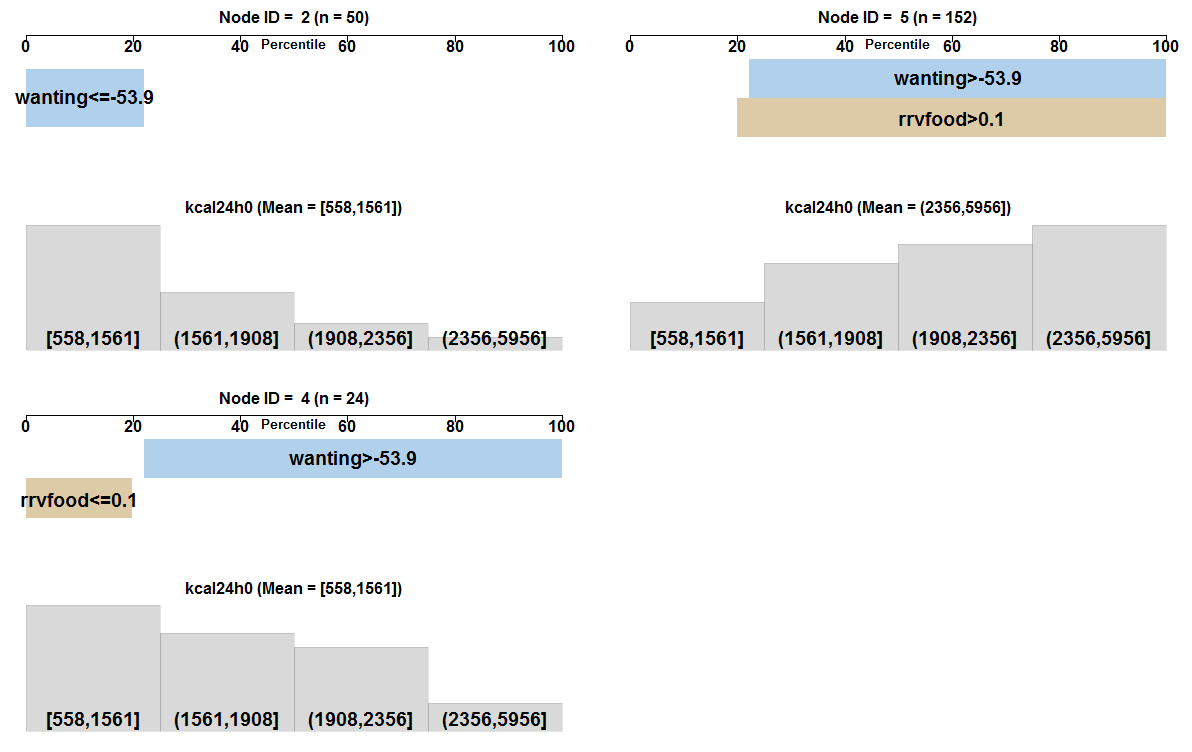}
\caption{The \texttt{visTree} function consolidates repeated splits on \textbf{skcal} in Figure \ref{fig:reptree} to display intervals which define subgroups.}\label{fig:catviz}
\end{figure}

\noindent By setting the argument \texttt{interval = TRUE}, the \texttt{visTree()} function can also display subgroups for trees with categorical outcomes. Figure \ref{fig:catviz} is a visualization of subgroups identified by the classification tree and the histogram plotted in each sub-plot has the category printed on each bin. 

\begin{verbatim}
visTree(ptree3, interval = T,  color.type = 1, alpha = 0.6)
\end{verbatim}

\subsection{Display controls}

\subsubsection{Color scheme and transparency}

The primary plotting function allows the user to modify the color scheme of horizontal colored bars that represent the splits at inner nodes and their corresponding criterions. A color palette can be selected using the option \texttt{color.type} and different types from the \texttt{colorspace} package can be specified in the plotting function using values 1 to 5. The type of color palettes include \texttt{rainbow\_hcl = 1}, \texttt{heat\_hcl = 2}, \texttt{terrain\_hcl = 3}, \texttt{sequential\_hcl = 4}, and \texttt{diverge\_hcl = 5}. The \texttt{visTree()} function uses the \texttt{rainbow\_hcl} as the default color palette, specified as \texttt{color.type = 1}. In addition, the \texttt{visTree} function also has the ability to control the level of transparency for the colored bars in the subplots using an option \texttt{alpha}. Values of \texttt{alpha} can be specified between 0 to 1 to modify the levels of transparency and are typically set to 0.5. 

\subsubsection{Text}

The \texttt{visTree()} function also offers options to modify the size of text for individual sub-plot titles, axes titles and axis labels placed on tick marks. These options include \texttt{text.title}, \texttt{text.axis}, \texttt{text.main}, \texttt{text.label}, \texttt{text.bar} and \texttt{text.percentile}. By default, these options are set as equal to 1.5 (except for \texttt{text.percentile} = 0.7). In addition, the horizontal colored bars are labeled by splits at inner nodes and their associated criterions and can be rounded to a specified number of decimal places. This is implemented within the \texttt{visTree()} function using \texttt{text.round} and by default is set to one decimal place \texttt{text.round = 1}. 

\subsubsection{Axes}

The \texttt{visTree()} function also allows the axes for the subplots to be modified as needed. For the axis associated with the values of the histogram (for continuous data) or the bar chart (for categorical data), the axis can be added using \texttt{add.h.axis = TRUE}. Similarly, the axis associated with percentile values for the horizontal bars can be added using \texttt{add.p.axis = TRUE}. 

\subsubsection{Density Curve}

For a continuous response, a density curve is overlaid on the histogram to represent the overall pattern. This density curve can be added or removed using a \texttt{density.line = TRUE} option in the \texttt{visTree()} function. By default, the \texttt{visTree()} function draws the density curve on the shaded histogram. 

\subsection{Implementation}

The \texttt{visTree()} function in the \texttt{visTree} package is developed utilizing the \texttt{`party'} object generated by the \texttt{ctree()} function within \texttt{partykit}. 
The main function which supports the creation of \texttt{visTree} plots is the \texttt{path\_node()} function, which traverses the tree and collects the set of nodes and associated split criterions that lead to each terminal node. For example, applied to the terminal node with node ID = 6, the function yields a string giving the splits along with the overall mean for that node. The plotting function \texttt{visTree()} parses the string to determine the population percentiles corresponding to each split and identify the subset of individuals belonging to this node. This function generates a series of sub-plots corresponding to each terminal node and the number of such sub-plots in the layout of the visualization is limited to 10. The size of decision trees generated by the \texttt{ctree()} function in \texttt{partykit} is controlled using \texttt{alpha} and the default parameter value set to 0.05 can be increased to grow larger decision trees. Fitting decision trees with a large number of terminal nodes to the \texttt{visTree()} function would result in a correspondingly large number of sub-plots. 

\subsection{Other decision tree types}

An advantage of leveraging the \texttt{partykit} structure is that \texttt{visTree} can easily be applied to other decision tree types (e.g., \texttt{`rpart'}) once they are converted to \texttt{`party'} objects.

\begin{verbatim}
library(rpart)
RP <- rpart(kcal24h0~liking+rrvfood, data = blsdata, 
        control = rpart.control(cp = 0.034))
partyRP <- as.party(RP)
partyRP
\end{verbatim}

\begin{verbatim}
Model formula:
kcal24h0 ~ liking + rrvfood

Fitted party:
[1] root
|   [2] rrvfood < 0.84444
|   |   [3] liking < -12.0625: 1660.134 (n = 78, err = 24027139.4)
|   |   [4] liking >= -12.0625: 2101.469 (n = 99, err = 29803292.7)
|   [5] rrvfood >= 0.84444: 2392.051 (n = 49, err = 34387536.3)

Number of inner nodes:    2
Number of terminal nodes: 3
\end{verbatim}

\noindent The output above represents a \texttt{`party'} object (obtained from a \texttt{`rpart'} object using the \texttt{rpart} package for a CART) that represents the relationship between two behavior covariates (\textbf{liking} and relative reinforcement of food (\textbf{rrvfood}) and daily caloric intake (in kilo-calories). This \texttt{`party'} object associated with the data set is applied to the \texttt{visTree} function. For example,   

\begin{verbatim}
visTree(RP, text.bar = 1.8, text.label = 1.4,
  text.round = 1, density.line = TRUE, text.percentile = 1.5)
\end{verbatim}

\begin{figure}[ht]
\centering
\includegraphics[scale = 0.45]{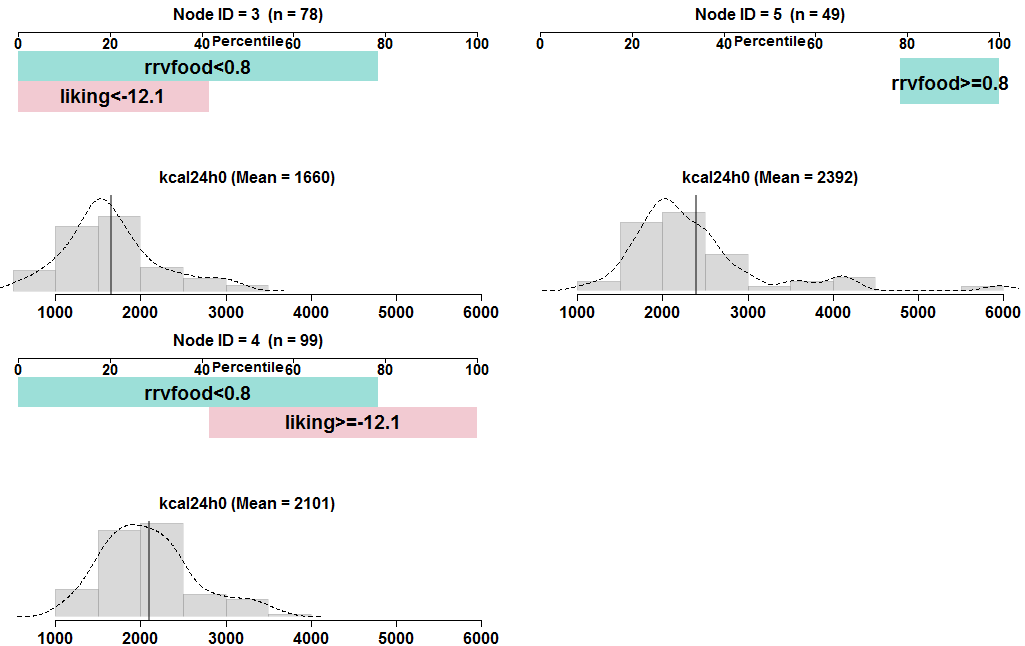}
\caption{Visualization to describe subgroups identified by a decision tree that was generated as a `rpart' object and coerced to a `party' object.}\label{fig:visrpart}
\end{figure}

\noindent Note that the plot in Figure \ref{fig:visrpart} uses a decision tree generated by \texttt{rpart()} which differs from the decision tree generated using \texttt{ctree()}. While the decision tree can be generated in similar style of output, a coerced tree does not have the statistical framework found in trees generated by the \texttt{ctree()} function. 
The \texttt{as.party} function can also coerce other trees such as the tree object generated by the \texttt{J48()} function in the \texttt{RWeka} package. The \texttt{J48()} function generates C4.5 decision trees and the coerced tree can be stored as a \texttt{`party'} object. 




\section{Summary}

The standard visualization of decision trees focuses on displaying individual splits and the distribution of outcomes within terminal nodes; it is not optimized for summarizing the population subgroups identified by the tree. The \texttt{visTree} package seeks to fill this gap by providing a graphical visualization tool which characterizes subgroups identified by decision trees. Using the toolkit provided by the \texttt{partykit} package, \texttt{visTree} provides functions to extract the underlying infrastructure to identify pathways to each defined subgroup. Our proposed visualization can be used with multiple decision tree types (e.g, CTree, CART, C4.5) and handles both categorical and continuous outcomes. In addition, the \texttt{visTree} package also accommodates multiple splits on a single covariate in the decision tree. This package is most useful for decision trees with a small to modest number of terminal nodes. A tree with a large number of terminal nodes (say $>$ 20) will produce a large number of \texttt{visTree()} subplots, which may be difficult to display on limited screen real estate. In practice, we recommend that prospective users of the \texttt{visTree} package restrict the number of terminal nodes in their tree by modifying options in their tree-building package. Lastly, while some customization options (e.g., controlling the color scheme and text size) are available, we are still working on the ability to exercise fine-grained control over the layout of the visualization and modify details within individual sub-plots.

\bibliography{main}
\bibliographystyle{plainnat}

\end{document}